# Project Based Learning of Embedded Systems

DANCO DAVCEV, BILJANA STOJKOSKA,
SLOBODAN KALAJDZISKI, KIRE TRIVODALIEV
Faculty of Electrical Engineering and Information Technologies
University "Sts. Cyril and Methodious"
Karpos 2, bb, 1000 Skopje
REPUBLIC OF MACEDONIA
etfdav@feit.ukim.edu.mk, biles@feit.ukim.edu.mk,
skalaj@feit.ukim.edu.mk, kire.trivodaliev@feit.ukim.edu.mk

*Abstract:* - Traditional teaching, usually based on lectures and tutorials fosters the idea of instruction-driven learning model where students are passive listeners. Besides this approach, Project Based Learning (PBL) as a different learning paradigm is standing behind constructivism learning theory, where learning from real-world situations is put on the first place.
The purpose of this paper is to present our approach in learning embedded systems at our University. It is based on combination of traditional (face-to-face) learning and PBL. Our PBL represents an interdisciplinary project based on wireless sensor monitoring of real-world environment (greenhouse). The students use UML that was shown as an excellent tool for developing such a projects. From the student perspective, we found that this high level of interdisciplinary is very valuable from the point of view of facing the students with real-life problems.

## 1 Introduction

An embedded system is a special-purpose system designed to perform one or a few pre-defined tasks, usually with very specific requirements. It can also be consider as an application that contains at least one programmable computer (microcontroller or microprocessor) and individuals that use it are unaware of its computer-based concepts [1]. Following these definitions, embedded systems range from very simple microcontrollers to very complex and sophisticated portable devices such as MP3 players, PDAs and smart phones [2][3].

Motivated by this, the goal of our undergraduate course of Embedded Systems was to enable the student to closely understand and become more familiar with the big diversity of the embedded "world". Our pedagogical approach in learning embedded systems is based on project-based learning methodology.

The project that should be designed and developed by the students that selected this course represents **an integration of different embedded devices (IDED)**. On one side we put small and smart sensors equipped with limited computational and wireless communication capabilities. These low-power and low-cost devices can be organized in a network and deployed widespread. On the other side, we have pocket PCs as complex type of embedded devices with more general-purpose. They have completely different power and bandwidth characteristics than low-power sensor nodes [4].

The IDED project is especially suitable for habitat and environmental monitoring, due to sensors ability to measure the environmental parameters and communicate among them and other smart devices and computers.

Section two of this paper describes the architecture of IDED. The third section explains the design and implementation of the IDED. The forth section gives the results provided from the tested participants in the project development. Section five presents the related work. Finally, we conclude this paper in section six.

## 2 IDED Architecture

In the recent years, there is a huge advancement in computing technology [5]. This has led to the production of wireless sensors that are not only capable of observing and reporting physical phenomena, but also, this new generation of tiny sensor nodes can accomplish other operations, like data processing and communication. The sensors are organized in a network and communicate and exchange information using radio modules. They are actually responsible for the first stages of the processing hierarchy [6]. After taking samples from the environment they sense (light level, air temperature, humidity etc.) they can process data or exchange it. All sent packets are collected by the sink node that is directly connected with the laptop.

Our project follows simple system architecture. All sensor nodes send their sensor readings to the base station, and the base station sends those data to users (clients) via the Internet. Pocket PCs represent the clients that communicate with the server. The medium through which the communication is accomplished is wireless local area network, i.e. the clients could be anywhere if they have connection to the Internet (Fig. 1).

The base station (Laptop or Desktop PC) serves as a link between the sensors and the Pocket PCs. It can simultaneously communicate to both the sensor network and the Pocket PCs.

In order to increase motivation and enthusiasm for the project among students, we considered that the project should be applicable in real-life. Thus the final product should be an application for monitoring and controlling a greenhouse.

The sensor nodes are deployed in a greenhouse room and are programmed to monitor different physical phenomena. They should especially pay attention for parameters that are important for providing normal conditions for the plants in the room. Sensors take samples on pre-defined intervals and send data to the sink node. Information is stored in a database and can be analyzed and processed on demand. If rapid or undesirable changes of some crucial values are noticed (i.e. if parameters exceed the predetermined range) system should alarm about the new situation for preventing or fixing the problem.

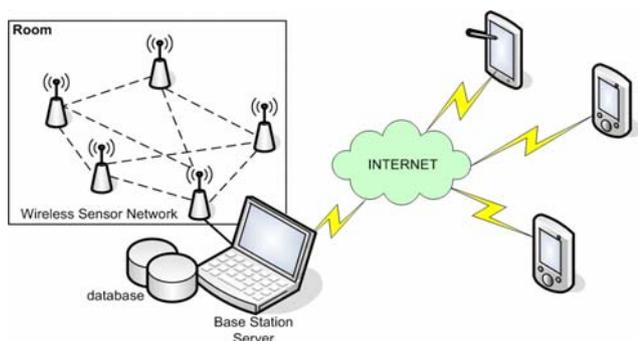

Fig. 1 Architecture of the system

Pocket PCs can be distributed anywhere within the system environment. In a real-life scenario they should be carried by the people that are employed in the greenhouse and are responsible for the garden and for the conditions inside. If system recognizes some alarming changes, the server should send an emergency signal to the clients (Pocket PCs), and employees should react according to that particular situation. Additionally, from time to time Pocket PC owner can ask for information about greenhouse conditions. The received data from the server can be processed locally and results can be shown on the display of the Pocket PC.

## 3 Design and Implementation

While UML (Unified Modelling Language) is an evolving standard for modelling and designing visual systems, it is rarely used for designing embedded systems [7].

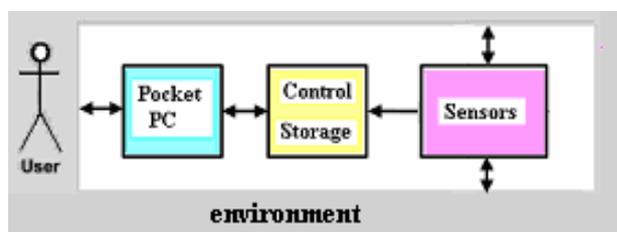

Fig. 2 Modified Broy logical architecture for IDED

We encourage our students to use UML and to implement object-oriented paradigm or to apply object analysis patterns in the analysis phase of embedded systems development [8].

As a start point in designing we use Broy logical architecture [8], [9] was used. Modified Broy logical architecture for IDED is shown on Fig. 2. IDED consists of five components: Control Center and Storage, Sensors, Pocket PC, the Environment, and the User. Interaction between the User and the Environment is accomplished via the Pocket PCs' user interface.

The deployment diagram of IDED can be seen in Fig. 3.

In IDED project, the students used Berkeley MICA motes, which can be purchased from Crossbow technology, Inc (see Fig.4). They are equipped with a radio module, 4 MHz microprocessor, memory, two AA batteries and a suite of sensors [10]. The motes run under TinyOS operating system, which is primary intended for embedded systems that have limitations and addresses its resources concurrency and resource management [4]. TinyOS, libraries and applications are all written in a new programming language named NesC. The sample taken from the sensor is represented as a 16-bit integer, but it contains a 10-bit ADC reading.

NesC is structured, component-based language. Due to its familiarity with the programming language C, it was not hard for the students to learn it easily. The application was first tested by using a simulator. After validating the simulation results, the application was deployed on the motes.

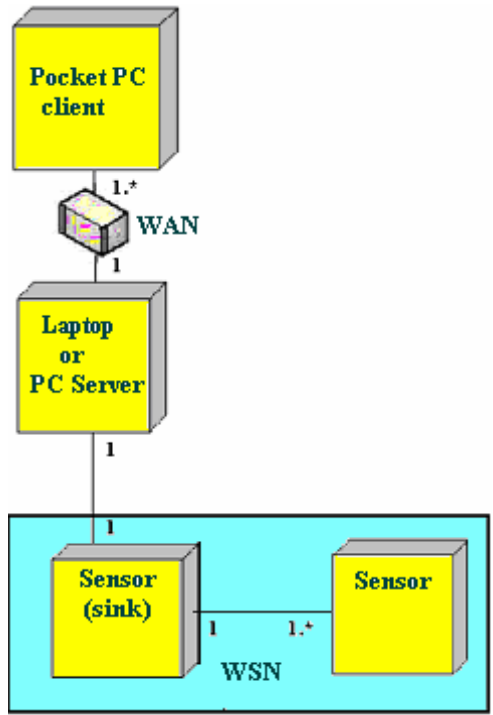

Fig. 3 Deployment diagram

Programming the motes can be done on two different ways, through serial port and using LAN. We have two sensor boards, MIB510 and MIB600, used for the different ways of communication. By doing this, students could experiment directly with hardware and see how things work "in the real world".

Because the motes are deployed in small and controlled environment we use a single-hop network was used. The major advantage of this system is its simple deployment [11].

The application for the pocket PCs is programmed in C# using Visual Studio .NET 2005, i.e. the Emulator for Pocket PC 2003, provided with the Visual Studio (see Fig.5).

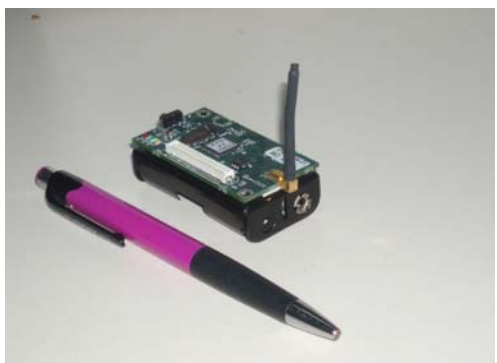

Fig. 4 Mica motes

The pocket PCs run under the Windows Mobile 2003 operating system, while the desktop machines, on which the application is practically deployed, run under Windows XP. The server uses PostgreSQL database for storing information and data received from the sensor nodes.

After testing the source code on the Emulator, the application was deployed on the Pocket PCs using ActiveSync, a synchronization program developed by Microsoft.

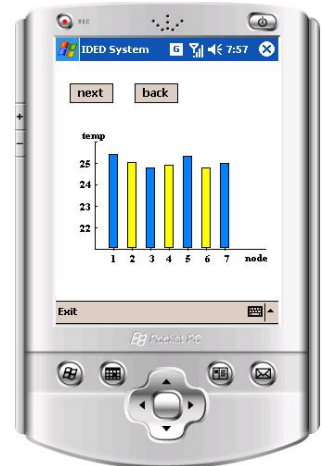

Fig. 5 Pocket PC Emulator

There are usually significant obstacles in setting up the application into the real hardware platform, mostly because testing on different embedded devices requires more expertise and working experience. Working directly with the hardware help students to learn all details of the embedded architecture, which is not possible if they only run their programs on simulators [3]. In our case, part of our laboratory was adopted as a garden room where wireless sensors were used by the students for monitoring process (see Fig.6).

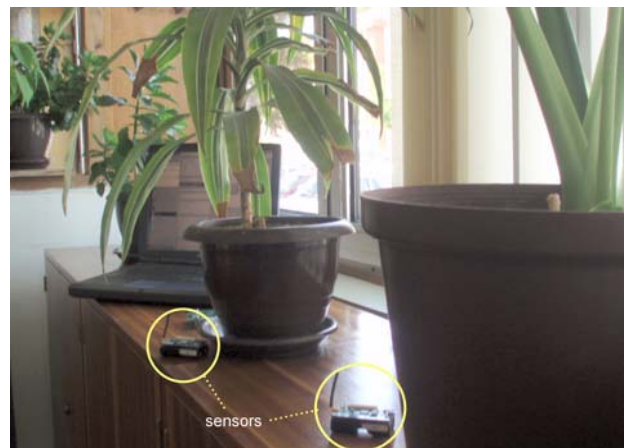

Fig. 6 Experimenting directly with hardware

We implemented PBL methodology as follows: the above described complex project is divided on three logical units, each of them presenting a separate mini project. The objective of the first mini project is to develop the application on the wireless sensor network (WSN), the objective of the second mini project is to build the application for Pocket PCs, and the third mini project is focused on the server side and data processing. Students are organized in groups of 3-4 students and each group is assigned one mini project. At the beginning students work together to analyze and design the hole project. In this phase the students can contact the professor. In the next stage, students work only within their own group and are focused on their particular task. After finishing the mini projects, groups have to work together again to integrate their mini systems into IDED. Working like this, students would collect specialized knowledge from one particular discipline, but they also have to obtain minimum knowledge about the other groups' mini projects. For example, the group developing the WSN project has to collect an enormous breadth of specific knowledge about WSNs [6], but in the process of integration this group has to understand the database solution from the mini project of the third group. While working together on the integration of the system, they are faced with a real-life situation when they have to demonstrate different skills like communication, collaboration, organization, resource management etc.

## 4  Evaluation

We tested many participants in the project development. Some of them were asked to fill a questionnaire with purpose to collect information and comments for the presented project based learning.

The questions list that examines participants experience and acceptability of the Project Based Learning are:

1) Do you think that Project Based Learning (PBL) makes the learning process easier?
2) Have you developed applications with similar purpose before?
3) Do you have previous working experience with smart sensors?
4) Do you have previous working experience with Pocket PC?
5) Do you find that PBL is more useful then traditional learning approach applied in similar courses (microcontrollers, wireless networks etc.)

6) Do you find that you needed previous knowledge and skills for efficient working on IDED?
7) Do you find that interdisciplinary projects increase the attractiveness of the course?

The Y axis in graph (1) represents questions with the same number from the questionnaire above.

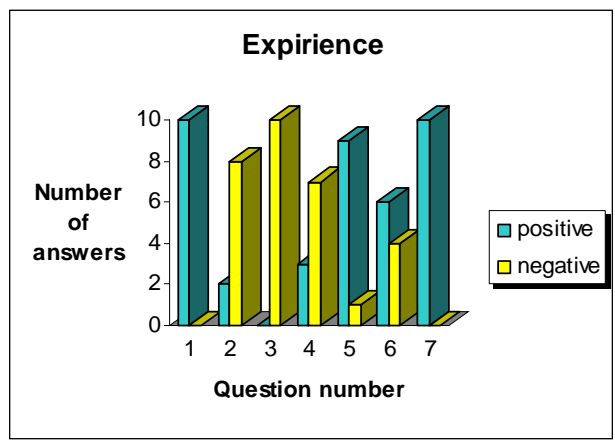

Graph (1) Results from the questionnaire.

Results show that some of the examinees have used and have previous experience with Pocket PC application, but none of them have worked with smart sensors. They think that previous knowledge helped them in developing the project, but it was not essential for the success of the project realization. All of them agreed that PBL could make learning proccess easier and is much more attractive than traditional learning.

General opinion of all examinees was that the project based learning can significantly increase the effectiveness of the learning process.

## 5  Related Work

As embedded systems are becoming more popular in both industry and the research groups, many universities began to offer courses in Embedded Systems. But the practice in teaching embedded systems is still considered to be outdated and not synchronized with the industry trends [3]. Traditional teaching, usually based on lectures and tutorials fosters the idea of instruction-driven learning model [12] where students are passive listeners. Besides this approach, PBL as a different learning paradigm is standing behind constructivism learning theory, where learning from real-world situations is put on the first place. According to this "student centered" strategy [13][14], students collect in-deeper knowledge when they directly experience the situation. The benefits from PBL are not only professional skills related to that particular project and specialized knowledge, but also include socio-technical aspects [14]. For successful development of the project, students should demonstrate different general skills, like ability to work in group, communication and collaboration, ability to monitor and periodically review the project, time management skills, leadership skills etc.

Hsu and Liu [15] describe a project based learning pilot course for embedded system education. They consider that in the success of the project development the most influential factor is previous learned knowledge and skills. But there is also effect from project subject selection and capability of working under time pressure.

In [13] a detailed description for designing Electronic Systems Curriculum using PBL strategy is given. The curriculum consists of four theoretical and four PBL courses. Theoretical courses should provide theoretical foundations for designing electronic circuits and should prepare students for the later PBL courses in which they should obtain multidisciplinary knowledge. Students are organized in groups and work together to develop complex projects. This approach has been evaluated for the last four years and provides magnificent results in increasing the interest in electronics among students. Multiple case studies to enhance PBL as a part of undergraduate computer architecture course are presented in [16]. Students and teachers simulate realistic consultant-customer scenario. Students play the roles of the consultants and they have to investigate some market sector and the existing computing technology. This is another alternative of PBL methodology.

In our approach, we added a PBL to the traditional face-to-face teaching by introducing an interdisciplinary project based on wireless sensor monitoring of real-world environment (greenhouse). The students use UML that was shown as an excellent tool for developing such a projects.. From the student perspective, we found that this high level of interdisciplinary is very valuable from the point of view of facing the students with real-life problems.

## 6   Conclusion

The purpose of this paper is to present our approach in learning embedded systems at our university. It is based on combination of traditional (face-to-face) learning and PBL. Our PBL represents an interdisciplinary project based on wireless sensor monitoring of real-world environment (greenhouse).

The students use UML that was shown as an excellent tool for developing such a projects. To the best of our knowledge we couldn't find in the literature a project with such a high level of interdisciplinary. From the student perspective, we found that this high level of interdisciplinary is very valuable from a point of view of facing the students with real-life problems.

In the future we plan to emphasize the PBL with respect to traditional learning for other similar courses. Interdisciplinary characteristics of the student projects will be especially respected.

Some of the suggestions and opinions that were given from the students and that were presented in our experimental results (for example the attractiveness of the course) will provide the base for the design of our future student projects.